\documentclass[twocolumn,showpacs,prl,floatfix]{revtex4}

\usepackage{graphicx}
\usepackage{dcolumn}
\usepackage{amsmath}

\usepackage{ae}

\begin{document}
%
%
\title[Spin Motion in Electron Transmission\ldots]{Spin Motion
  in Electron Transmission through Ultrathin Ferromagnetic Films\\
  Accessed by Photoelectron Spectroscopy}

\author{J. Henk}
\email[Corresponding author. Electronic address:\ ]{henk@mpi-halle.de}
\author{P. Bose}
\author{Th.\ Michael}
\author{P. Bruno}
\affiliation{Max-Planck-Institut f\"ur Mikrostrukturphysik, Weinberg~2,
  D-06120 Halle (Saale), Germany}

\date{\today}

\begin{abstract}
  \textit{Ab initio} and model calculations demonstrate that the spin
  motion of electrons transmitted through ferromagnetic films can be
  analyzed in detail by means of angle- and spin-resolved core-level
  photoelectron spectroscopy.  The spin motion appears as precession
  of the photoelectron spin polarization around and as relaxation
  towards the magnetization direction.  In a systematic study for
  ultrathin Fe films on Pd(001) we elucidate its dependence on the Fe
  film thickness and on the Fe electronic structure.  In addition to
  elastic and inelastic scattering, the effect of band gaps on the
  spin motion is addressed in particular.
\end{abstract}

\pacs{75.70-i, 79.60.-i, 73.40.Gk, 75.50.Bb}

\maketitle

%
%
To take advantage of the spin in electronic devices, in order to form
new ``spintronic'' devices, is currently in progress worldwide.  This
goal challenges both applied and basic physics, the latter being
mostly concerned with model systems of spin-dependent
transport~\cite{Maekawa02}. Aiming at very small devices, properties
of magnetic nanostructures become increasingly important. In
particular, spin-dependent scattering in ultrathin films and at
interfaces may have a profound effect on the transport
properties~\cite{Parkin96,Zahn98b}: the electronic spins start to
precess and the spin current applies a spin-transfer torque on the
magnetization in the ferromagnet.  To understand in detail the spin
motion in electron transmission through magnetic films, one obviously
needs a microscopic probe.

Ferromagnetic resonance, used successfully to study magnetic
properties of multilayer systems~\cite{Bland94}, unfortunately cannot
deal with electron transmission.  However, spin- and time-resolved
photoelectron spectroscopy was employed to investigate directly spin
filtering in the time domain~\cite{Aeschlimann98}.  Another successful
method is transmission of spin-polarized electrons (usually produced
with a GaAs source) through freestanding ferromagnetic
films~\cite{Lassailly94}.  This way, the spin motion which shows up as
precession of the electron spin-polarization (ESP) $\vec{P}$ around
the magnetization direction and as relaxation of $\vec{P}$ towards the
magnetization $\vec{M}$ was investigated~\cite{Oberli98,Weber99}.

\begin{figure}
  \begin{center}
    \includegraphics[scale = 0.95]{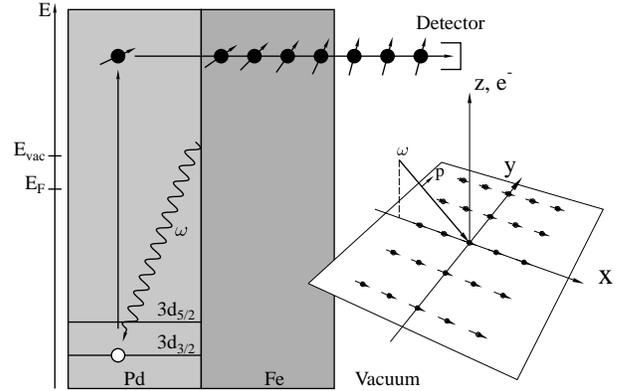}
    \caption{Spin motion in electron transmission through
      a ferromagnetic film accessed by photoelectron spectroscopy.
      Left: A core electron is excited by the incident radiation (wavy
      line, photon energy $\omega$) in the Pd substrate (light grey).
      The spin polarization (arrow) of the photoelectron (filled
      circle) is oriented due to spin-orbit interaction. During the
      transmission through the magnetic Fe film (dark grey), the spin
      polarization rotates further (spin motion) but stops rotating in
      the vacuum.  $E_{\mathrm{F}}$ and $E_{\mathrm{vac}}$ are the
      Fermi and the vacuum level, respectively. Right: Set-up of
      normal photoemission from a ferromagnetic surface (with
      magnetization along $x$) and p-polarized light incident in the
      $xz$ plane (photon energy $\omega$).}
    \label{fig:geometry}
  \end{center}
\end{figure}

In this Letter, we propose to apply angle- and spin-resolved
photoelectron spectroscopy from core levels to access directly the
spin motion of electrons transmitted through an ultrathin
ferromagnetic film (Fig.~\ref{fig:geometry}).
The ``theoretical experiments'' reported here rely in particular on
the possibility to orient the spin polarization of the incoming
photoelectrons by the incident light, an effect which is due to
spin-orbit coupling.  In the following, the basic ideas of our
approach are described for the chosen systems $n$~ML Fe/Pd(001), $n =
1, \ldots, 6$ (for details, see Ref.~\cite{Henk02a}).

(i)~The incident light excites electrons from Pd-$3d_{3/2}$ core
levels of the Pd(001) substrate into a state above the vacuum level
$E_{\mathrm{vac}}$.

(ii)~Choosing linearly p-polarized light with incidence direction
given by $\vartheta_{\mathrm{ph}} = 45^{\circ}$ polar angle and
variable azimuth $\varphi_{\mathrm{ph}}$, the photoelectron spin in
the substrate can be aligned to any desired direction in the $xy$
surface plane (Cartesian coordinates are defined in
Fig.~\ref{fig:geometry}). It was theoretically and experimentally
shown for nonmagnetic layered systems with fourfold rotational
symmetry that an ESP perpendicular to the scattering plane (spanned by
the surface normal and the incidence direction; see
Ref.~\cite{Henk97a} and refs.\ therein) is produced:
$\vec{P}^{\mathrm{in}} \propto (-\sin\varphi_{\mathrm{ph}},
\cos\varphi_{\mathrm{ph}}, 0)$.  For $\varphi_{\mathrm{ph}} =
0^{\circ}$ and $180^{\circ}$, $\vec{P}^{\mathrm{in}}$ is perpendicular
to the magnetization $\vec{M}$ (which is parallel to $x$).  Hence, the
commonly used external GaAs source for spin-polarized electrons is, so
to speak, replaced by an internal one, with the advantage of easy
orientation of $\vec{P}^{\mathrm{in}}$. Choosing other light
polarizations and incidence directions one can produce
$\vec{P}^{\mathrm{in}}$ with a component along the surface
normal~\cite{Henk96a}.

(iii)~During the transmission through the Fe film, the photoelectron
is subject to elastic and inelastic processes. Both can simply be
modeled by spin-dependent scattering at an asymmetric quantum well
which comprises the substrate-film and the film-vacuum interfaces. The
transmitted ESP $\vec{P}^{\mathrm{tr}}$ reads
\begin{equation}
  \label{eq:ptr}
  \vec{P}^{\mathrm{tr}}
  \propto
  \left(
    \begin{array}{c}
      |T^{\uparrow}|^{2} - |T^{\downarrow}|^{2} + P_{x}^{\mathrm{in}} \left(|T^{\uparrow}|^{2} + |T^{\downarrow}|^{2}\right)
      \\
      P_{y}^{\mathrm{in}} \mathrm{Re}({T^{\uparrow}}^{\star} T^{\downarrow})
      -
      P_{z}^{\mathrm{in}} \mathrm{Im}({T^{\uparrow}}^{\star} T^{\downarrow})
      \\
      P_{z}^{\mathrm{in}} \mathrm{Re}({T^{\uparrow}}^{\star} T^{\downarrow})
      +
      P_{y}^{\mathrm{in}} \mathrm{Im}({T^{\uparrow}}^{\star} T^{\downarrow})
    \end{array}
  \right),
\end{equation}
where the spin-dependent transmission coefficients
$T^{\uparrow(\downarrow)}$ take into account multiple scattering.
Considering elastic scattering only, the dependence of
$\vec{P}^{\mathrm{tr}}$ on the film thickness $d$ shows two typical
oscillation periods that depend on the electron wavenumbers
$k_{z}^{\uparrow(\downarrow)}$ in the film. The longer period with
wavelength $2 \pi / (k_{z}^{\uparrow} - k_{z}^{\downarrow})$ describes
the precession of the transversal components $P_{y}^{\mathrm{tr}}$ and
$P_{z}^{\mathrm{tr}}$ around $\vec{M}$~\cite{Stiles02a,Stiles02c}.
The short-period oscillation with wavelength $2 \pi /
(k_{z}^{\uparrow} + k_{z}^{\downarrow})$ and much smaller amplitude is
due to multiple scattering at the interfaces. The longitudinal
component $P_{x}^{\mathrm{tr}}$ remains constant on average.

Inelastic scattering leads to spin-dependent attenuation within the
film. This can simply be simulated by multiplying the phase factors
that describe the propagation between the interfaces by $\exp(-d /
\lambda^{\uparrow(\downarrow)})$.  This spin-filter effect relaxes
$\vec{P}^{\mathrm{tr}}$ towards $\vec{M}$ (i.\,e., $\lim_{d \to
  \infty} P_{x}^{\mathrm{tr}} = 1$ for $\lambda^{\uparrow} >
\lambda^{\downarrow}$).  It was successfully used to determine the
attenuation lengths
$\lambda^{\uparrow(\downarrow)}$~\cite{Pappas91,Gokhale91} and to
obtain the spin-resolved electronic structure of Fe~\cite{Kuch95a}.
There is no spin motion in nonmagnetic regions (e.\,g., vacuum).

(iv)~The photoelectrons are eventually detected spin-resolved in
normal emission ($\vec{k}_{\parallel} = 0$).  Note that the electron
energies are considerably larger than those in spin-dependent
transport measurements.

The small photoelectron escape depth~\cite{Seah79,Jablonski02}
restricts $d$ to a few ML\@. This implies for ultrathin films that the
short-period oscillation might dominate the spin motion, a complete
precession cannot be observed, and the relaxation limit ($\vec{P}
\parallel \vec{M}$) cannot be reached in practice.

\paragraph{Theoretical.}
Starting from first-principles electronic-structure calculations for
0--6 ML fcc-Fe/Pd(001) (local spin-density approximation of
density-functional theory; screened KKR method; for details, see
Ref.~\cite{Henk02a}) we computed spin- and angle-resolved
constant-initial state photoemission spectra within the relativistic
one-step model, as formulated in the layer-KKR method (see, e.\,g.,
Ref.~\cite{Henk01c}).

We choose Fe/Pd(001) due to the large magnetic moment of Fe and the
strong spin-orbit coupling in Pd which results in a sizable
$\vec{P}^{\mathrm{in}}$. The covering Fe induces a magnetic moment of
about $0.24~\mu_{\mathrm{B}}$ in the Pd layer close to the Fe/Pd
interface~\cite{Henk02a}. Hence, $\vec{P}^{\mathrm{in}}$ originates
from the induced exchange splitting and from spin-orbit coupling.  To
reveal the origin of the spin motion, we considered various
``artificial'' magnetic configurations in which the magnetization in
the Fe film or in the Pd substrate were switched off separately.
Further, changing the inverse photoelectron lifetime in the Fe film
allows to differentiate between elastic (precession around $\vec{M}$)
and inelastic processes (relaxation towards $\vec{M}$).

\begin{figure}
  \begin{center}
    \includegraphics[scale = 0.95]{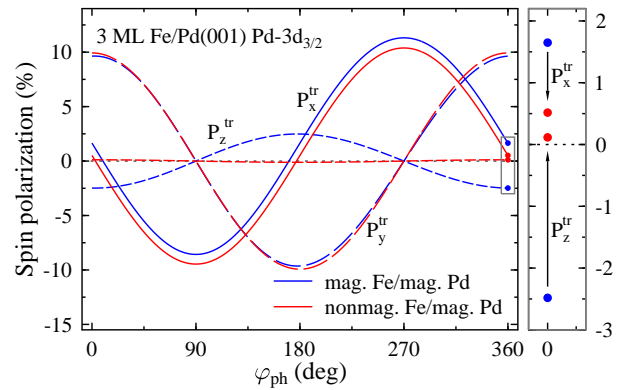}
    \caption{Transmitted spin polarization $\vec{P}^{\mathrm{tr}}$ versus
      azimuth $\varphi_{\mathrm{ph}}$ of light incidence for 3~ML
      Fe/Pd(001) and $27~\mathrm{eV}$ kinetic energy. Blue (red) lines
      are for magnetic (nonmagnetic) Fe on magnetic Pd (see text). The
      right panel shows data for $\varphi_{\mathrm{ph}} = 0^{\circ}$
      or $360^{\circ}$ enlarged (cf.\ the grey rectangle).}
    \label{fig:filmablge}
  \end{center}
\end{figure}

\paragraph{Existence and origin of the spin motion.}
Considering the configuration with all Fe layers and the Pd layers
close to the Fe/Pd interface being magnetic (``mag.\ Fe/mag.\ Pd'',
blue in Fig.~\ref{fig:filmablge}) and $\vec{P}^{\mathrm{in}}$ along
the $y$ axis ($\varphi_{\mathrm{ph}} = 0^{\circ}$; right panel), the
existence of the spin motion is established by nonzero
$P_{x}^{\mathrm{tr}}$ and $P_{z}^{\mathrm{tr}}$.
For $\varphi_{\mathrm{ph}} = 90^{\circ}$ and $270^{\circ}$ (i.\,e.,
$\vec{P}^{\mathrm{in}} \parallel \vec{M}$) both $P_{y}^{\mathrm{tr}}$
and $P_{z}^{\mathrm{tr}}$ vanish, leaving only $P_{x}^{\mathrm{tr}}$
nonzero in agreement with Eq.~(\ref{eq:ptr}).  The dependence of
$\vec{P}^{\mathrm{tr}}$ on $\varphi_{\mathrm{ph}}$ (left panel)
follows perfectly that derived analytically in Ref.~\cite{Henk96a}.

Setting all Fe layers artificially nonmagnetic but keeping the
magnetism in the Pd interface layers (``nonmag.\ Fe/mag.\ Pd'', red in
Fig.~\ref{fig:filmablge}), $P_{z}^{\mathrm{tr}}$ is strongly reduced
(in absolute value).  This confirms that the spin motion originates
dominantly from the magnetism in the Fe film.  In the complete
nonmagnetic configuration (not shown), $P_{z}^{\mathrm{tr}}$ vanishes.

\paragraph{Elastic and inelastic processes.}
Inelastic processes can be simulated in calculations by adding an
imaginary self-energy to the potential (see, e.\,g.,
Ref.~\cite{Rundgren99}). To unveil the influence of these processes,
we reduced the inverse photoelectron lifetime in the Fe film to
$0.001~\mathrm{eV}$ (``elastic'' case), compared to the otherwise
chosen $1.8~\mathrm{eV}$ (``inelastic'' case).
\begin{figure}
  \begin{center}
    \includegraphics[scale = 0.95]{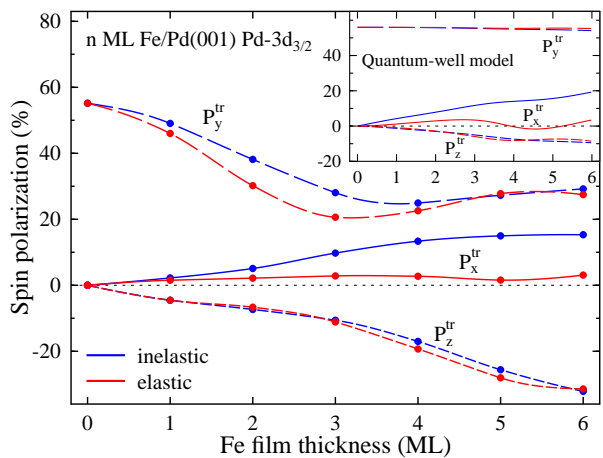}
    \caption{Elastic and inelastic effects in electron transmission 
      through 0--6 ML Fe on Pd(001) at $17.5~\mathrm{eV}$ kinetic
      energy and azimuth of light incidence $\varphi_{\mathrm{ph}} =
      0^{\circ}$. The transmitted electron-spin polarization
      $\vec{P}^{\mathrm{tr}}$ is shown versus Fe-film thickness $n$
      (in ML) for the ``inelastic'' (blue) and the ``elastic'' (red)
      case. The inset shows corresponding results of a model
      calculation.}
    \label{fig:filmpolb}
  \end{center}
\end{figure}
Being rather small and almost constant in the ``elastic'' case,
$P_{x}^{\mathrm{tr}}$ increases with Fe coverage in the ``inelastic''
case (Fig.~\ref{fig:filmpolb}), i.\,e., $\vec{P}^{\mathrm{tr}}$ starts
to relax towards $\vec{M}$.  Because the short-period oscillation is
relevant for ultra-thin films, the precession of
$\vec{P}^{\mathrm{tr}}$ around $\vec{M}$ (which shows the long
wavelength) cannot be clearly observed.  To corroborate these
findings, we calculated $\vec{P}^{\mathrm{tr}}$ within the
quantum-well model sketched preceding, with parameters obtained from
the Pd and Fe bulk-band structures (inset in Fig.~\ref{fig:filmpolb}).
The resulting wavelengths of about 200~ML (precession) and 3.9~ML
(multiple-scattering) lead to reasonable agreement concerning
$P_{x}^{\mathrm{tr}}$ and $P_{z}^{\mathrm{tr}}$.  However,
$P_{y}^{\mathrm{tr}}$ does not show such pronounced a minimum at 3--4
ML\@. The differences between model and \textit{ab initio}
calculations can be attributed to the number of transmission channels:
a single one in the model but several channels (with different
wavelengths) in the \textit{ab initio} calculations.

\paragraph{Effects of the electronic structure.}
To show how the spin motion depends on details of the electronic
structure, we address spin-resolved constant-initial-state
photoemission spectra. In contrast to SPLEED experiments in which
$\vec{P}^{\mathrm{in}}$ is typically parallel or antiparallel to the
magnetization~\cite{Scheunemann97b,Egger99}, we choose a transverse
$\vec{P}^{\mathrm{in}}$ ($\varphi_{\mathrm{ph}} = 0^{\circ}$).  For
clarity reasons, the following discussion rests upon the complex bulk
band structure, rather than on layer- and spin-resolved spectral
densities.

\begin{figure}
  \begin{center}
    \includegraphics[scale = 0.95]{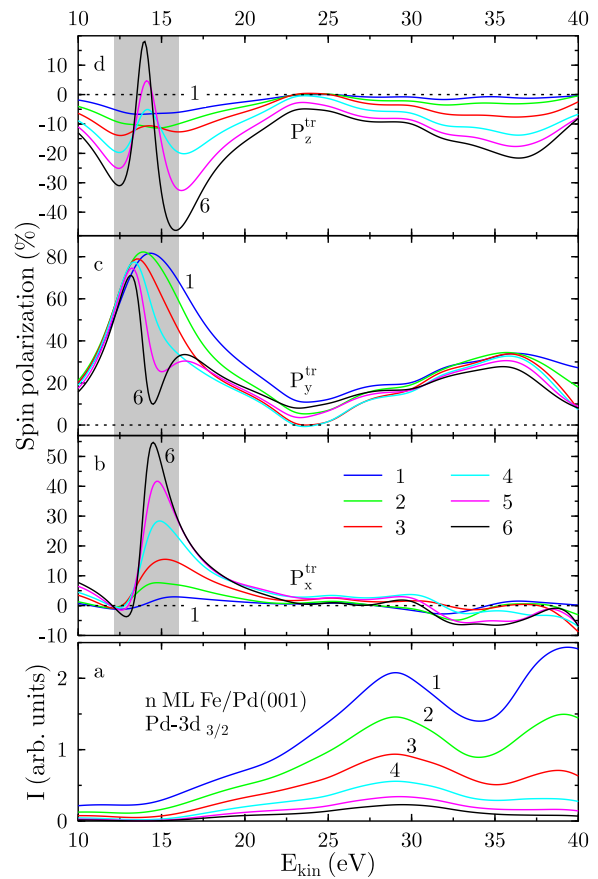}
    \caption{Energy dependence of the spin motion for
      1--6 ML Fe on Pd(001). (a) Spin-averaged constant-initial-state
      photoemission intensities $I$ versus kinetic energy
      $E_{\mathrm{kin}}$ of the photoelectrons.  (b)--(d) Transmitted
      electron-spin polarization $\vec{P}^{\mathrm{tr}}$. The Fe-film
      thickness $n$ (in ML) is indicated by numbers and color-coding.
      The grey area highlights a prominent feature discussed in the
      text.}
    \label{fig:cis.p}
  \end{center}
\end{figure}

The spin-averaged intensities (Fig.~\ref{fig:cis.p}a) decrease
significantly with Fe coverage, caused by the small photoelectron
escape depth. The global shape of the spectra, however, remains almost
unaffected.  Changes of the slopes, best to be seen for 1~ML Fe but
present for all Fe-film thicknesses, can be traced back to the Fe
electronic structure (not shown): an increase of the slope is
associated with the onset of additional transmission channels, i.\,e.,
dispersive Fe bands.  In particular, one pair of spin-split bands
provides efficient transmission channels, which leads to the intensity
increase at about $15~\mathrm{eV}$.  The pronounced minimum at about
$34~\mathrm{eV}$ kinetic energy can be attributed to a Pd-band gap,
which reduces the number of transmission channels in the substrate.

At higher energies where several transmission channels contribute, the
evolution of $\vec{P}^{\mathrm{tr}}$ with Fe coverage is rather
complicated (Figs.~\ref{fig:cis.p}b--d). Therefore, we concentrate on
low energies where the number of channels is small and the evolution
is almost monotonous. The most significant structures that increase
with Fe coverage show up between $12~\mathrm{eV}$ and $16~\mathrm{eV}$
(grey area in Figs.~\ref{fig:cis.p}b--d): $P_{x}^{\mathrm{tr}}$ and
$P_{y}^{\mathrm{tr}}$ display $-/+$ and $+/-$ modulations, resp.,
accompanied by a maximum in $P_{z}^{\mathrm{tr}}$.  A detailed
analysis corroborates its relation to the Fe electronic structure and
shows that it can be attributed to exchange-split band gaps in
conjunction with the onset of additional transmission channels in that
particular energy range.

\begin{figure}
  \begin{center}
    \includegraphics[scale = 0.95]{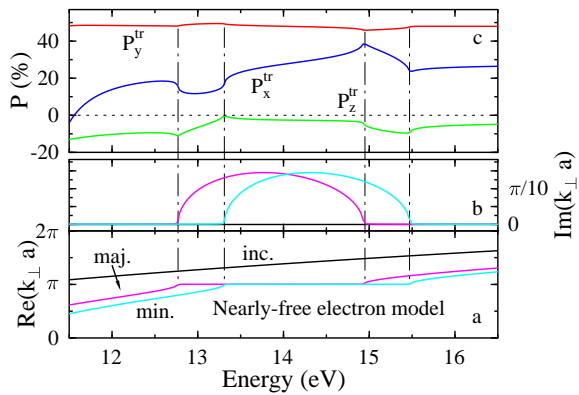}
    \caption{Effect of exchange-split band gaps in the film electronic
      structure on the spin motion.  (a) and (b): Complex band
      structure of the substrate (dash-dotted, ``inc.'') and the
      magnetic film (majority, solid, ``maj.''; minority, dashed,
      ``min.'') in the extended zone scheme. (c) Electron spin
      polarization of the transmitted electrons. For details, see
      text. Vertical dash-dotted lines serve as guides to the eye.}
    \label{fig:transmission.model}
  \end{center}
\end{figure}

To provide direct evidence that band gaps manifest themselves in
pronounced spin-motion structures, we calculated the ESP in an
inelastic three-band nearly-free-electron model. The substrate is
taken as semi-infinite free space (with zero potential), whereas a
nonzero scattering potential in the magnetic film gives rise to
exchange-split band gaps (Figs.~\ref{fig:transmission.model}a and b).
For $P_{y}^{\mathrm{in}} = 50\%$ in a representative set-up,
$P_{x}^{\mathrm{tr}}$ and $P_{y}^{\mathrm{tr}}$ show a $-/+$ and a
small $+/-$ modulation, resp., whereas $P_{z}^{\mathrm{tr}}$ increases
in the band-gap middle (Fig.~\ref{fig:transmission.model}c). These
findings can be explained by the reduced transmission of one spin
channel which is mediated by evanescent states [nonzero
$\mathrm{Im}(k_{z})$ in Fig.~\ref{fig:transmission.model}b; note that
transverse spinors are given by a weighted sum of spin-up (``maj.'')
and spin-down (``min.'') Pauli spinors].  Although the Fe band
structure is much more complicated, the structures in the model
calculation have counterparts in Figs.~\ref{fig:cis.p}b--d (grey
area). That distinct band-gap related features do not show up in
Fig.~\ref{fig:cis.p} at higher kinetic energies can be attributed to
the onset of efficient transmission channels just at about
$15~\mathrm{eV}$.

\paragraph{Conclusions.}
We have shown by means of ``theoretical experiments'' that the spin
motion in electron transmission through ferromagnetic films can be
analyzed in detail by angle- and spin-resolved photoelectron
spectroscopy. As main advantages of this approach one might consider
that the preparation of freestanding films is avoided and that the
spin polarization of the incoming electrons can easily be oriented.
Since intensities and spin polarizations depend significantly on the
film thickness, one obtains information on the electronic structure,
in particular on that of the film. Realistic calculations for Fe films
on Pd(001), that are to be confirmed experimentally, suggest promising
analyses of spin-dependent transport through magnetic layers. Further,
we regard our approach as a tool for investigations of magnetic
configurations, with the possibility of analyzing noncollinear spin
structures.

%
%
\bibliography{short,refs,refs2,refs3}
\bibliographystyle{apsrev}

\end{document}